\documentclass[
prd,
superscriptaddress,
10 pt,
twocolumn,
preprintnumbers,
notitlepage,
secnumarabic,
nobibnotes,
nofootinbib,
showpacs
]{revtex4-1}

\usepackage[T1]{fontenc}
\usepackage[pdftex]{color,graphicx}
\usepackage{mathtools}
\usepackage{amssymb}
\usepackage{bm}
\usepackage{dsfont}
\usepackage{mathrsfs}
\usepackage{calligra}
\usepackage{tabularx}
\usepackage[artemisia]{textgreek}
\usepackage{hyperref}
\usepackage{xcolor}
\usepackage{enumerate}
\usepackage{multirow}
\usepackage{url}
\usepackage{caption}
\usepackage[normalem]{ulem}
\usepackage{perpage}
\MakePerPage{footnote}
\renewcommand{\.}{\!\;}		%	2 point space
\renewcommand{\@}{\!\:\!}	%	-2 point space

\DeclareMathAlphabet\mathbfcal{OMS}{cmsy}{b}{n}

\renewcommand{\i}{\!\:\mathrm{i}}

\begin{document}
\title{Lie algebra of Ashtekar-Barbero connection operators}

\author{Jakub Bilski}
\email{bilski@zjut.edu.cn}
\affiliation{Institute for Theoretical Physics and Cosmology, Zhejiang University of Technology, 310023 Hangzhou, China}

	%%%%%%%%%	%%%%%%%%%	%%%%%%%%%	%%%%%%%%%

	%%%%%%%%%	%%%%%%%%%	%%%%%%%%%	%%%%%%%%%
\begin{abstract}
\noindent
Holonomies of the Ashtekar-Barbero connection can be considered as abstract elements of a Lie group exponentially mapped from their connections representation. This idea provides a possibility to compare the geometric and algebraic properties of these objects. The result allows to identify the next-to-the-leading-order terms in the geometric and algebraic expansion of a holonomy. This identification leads to the verification of the related Hilbert space formulation. If states are the representations of the holonomy's symmetry group, they preserve gauge transformations according to Wigner's theorem. Thus, the spin network in loop quantum gravity satisfies this theorem. Moreover, the considered identification of the different expansions ensures the reality of the Ashtekar connection. Only the holonomies of real connections lead to the formulation of states that satisfy Wigner's theorem.
\end{abstract}
	%%%%%%%%%	%%%%%%%%%	%%%%%%%%%	%%%%%%%%%

\maketitle

	%%%%%%%%%	%%%%%%%%%	%%%%%%%%%	%%%%%%%%% 

	%%%%%%%%%	%%%%%%%%%	%%%%%%%%%	%%%%%%%%%
\section{Motivation}\label{Sec_Motivation}

\noindent
The Ashtekar variables \cite{Ashtekar:1986yd} are the representation of the gravitational degrees of freedom and allow to express a particular action \cite{Holst:1995pc}, leading to the Einstein field equations \cite{Einstein:1916vd} in a form of a gauge theory, analogous to the Yang-Mills model \cite{Yang:1954ek}. In the case of the real Ashtekar variables, they lead to the Hamiltonian \cite{Holst:1995pc,Barbero:1994ap}, which may be considered as a candidate for the classical limit of the background-independent formulation of quantum gravity called loop quantum gravity (LQG) \cite{Ashtekar:2004eh,Thiemann:1996aw}. It is worth bringing up that the complex Ashtekar variables would lead to rather nonphysical candidates for observables --- see the expansion in \eqref{Wigner_expansion}, in which all parameters have to be real. In this case, auxiliary effective methods \cite{Barbero:1994an,Thiemann:1995ug} of reality conditions implementation on the classical phase space would be needed.

It is known that the symmetry transformations of operators, which preserve probabilities at the quantum level, are related to a Lie algebra --- \textit{cf.} \cite{Weinberg:1995mt}. The same algebra describes the symmetries of the candidates for field theory operators. We are going to analyze the particular candidates forming the canonical pair for the background-independent tetrad-based models of quantum gravity with a time gauge \cite{Arnowitt:1960es} (mainly most variants of LQG). These variables are the real Ashtekar-Barbero connection and the densitized dreibein \cite{Barbero:1994ap,Ashtekar:2004eh,Thiemann:2007zz}. However, they are not directly quantized but require rather special procedures of regularization \cite{Thiemann:1996aw,Thiemann:2007zz}, in which the Ashtekar-Barbero connection is replaced by a holonomy --- see \eqref{holonomy_general} and \eqref{holonomy_expansion}.

Apart from the reality of the Lie algebra-valued variables, it is required that the related generators are either unitary and linear or antiunitary and antilinear \cite{Wigner:1931}. As in the case of the standard operators in quantum mechanics and quantum field theory, they are unitary and linear in general \cite{Weinberg:1995mt}. All of the aforementioned facts encouraged this investigation of the holonomy representation algebra.

	%%%%%%%%%	%%%%%%%%%	%%%%%%%%%	%%%%%%%%% 

	%%%%%%%%%	%%%%%%%%%	%%%%%%%%%	%%%%%%%%%
\section{Holonomy of the Ashtekar-Barbero connection}\label{Sec_Holonomy}

\noindent
The parallel transport of a vector bundle element over the manifold $M$ to another bundle along a smooth oriented path $\ell(s):[0,1]\to M$ is determined by the expression
\begin{align}
\label{holonomy_general}
h_{[0,1]}^{-1}[\bm{A}]=\mathcal{P}\exp\!\bigg(\!-\!\int_{\@0}^1\!\!\!\!ds\,\dot{\ell}^a\@(s)\.\bm{A}_a\@\big(\ell(s)\big)\!\bigg).
\end{align}
This object is called the holonomy of the vector potential $\bm{A}:=A^I\bm{t}_I$ along the path $\ell$, where $\bm{t}_I$ are the generators of a Lie algebra. The `inverse notation', $h^{-1}$, follows the convention present in the LQG-related literature \cite{Ashtekar:2004eh,Thiemann:2007zz}. Consequently, $h$ denotes the inverse holonomy. Taking a piecewise smooth path $\mathscr{L}:=\ell^1\circ\ell^2\circ...$, the holonomy is defined as a composite functional of smooth pieces $\ell^1,\ell^2,...\.$. The Ashtekar-Barbero connection coefficients $A^I\in\mathds{R}$ are specified by the following normalization of the Lie bracket,
\begin{align}
\label{Lie_real}
[\bm{t}_J,\bm{t}_K]=\frac{1}{2}C^I_{\ JK}\bm{t}_I\,,
\end{align}
where $C^I_{\ JK}=-C^I_{\ KJ}\in\mathds{R}$ are the structure constants. By definition, this normalization resolves the issue of reality conditions implementation.

The formalism of LQG (\textit{cf.} \cite{Ashtekar:2004eh,Thiemann:2007zz}) introduces a graph structure, allowing to define the associated scalar product as a relation between the positions of variables on this graph, hence without using a metric. This technique is based on the framework constructed in \cite{Ashtekar:1994mh,Ashtekar:1994wa,Ashtekar:1995zh}. The holonomy  $h_p^{-1}=\mathcal{P}\exp\!\big(-\@\int_0^{l_p}\@\bm{A}\big)$, adjusted to a particular graph's edge (of length $l_p:=\mathbb{L}_0\varepsilon_p$) can be expanded around the infinitesimal value of the dimensionless regularization parameter $\varepsilon_p:=\varepsilon_{l_p}\in(0,1)$,
\begin{align}
\begin{split}
\label{holonomy_expansion}
h_{(\@p\@)}^{\mp1}[\bm{A}]=&\;\mathds{1}\mp\mathbb{L}_0\varepsilon_{(\@p\@)}\bm{A}_{(\@p\@)}
+\frac{1}{2}(\mathbb{L}_0\varepsilon_{(\@p\@)}\@)^2\bm{A}_{(\@p\@)}\bm{A}_{(\@p\@)}
\\
&\mp\frac{1}{2}(\mathbb{L}_0\varepsilon_{(\@p\@)}\@)^2\.\partial_{(\@p\@)}\bm{A}_{(\@p\@)}
+\mathcal{O}(\varepsilon^3)\,.
\end{split}
\end{align}
The indices written in the brackets are not summed and $\mathbb{L}_0$ is a fiducial length scale. It is worth emphasizing that all the elements are oriented in the same direction. It is the result of the fixed orientation of the line $\dot{\ell}_{(\@p\@)}\@(s)$ in \eqref{holonomy_general} tangent to $l_p$ at the initial point of the edge, namely at $s=0$. This tangent line determines a one-dimensional system and defines the spatial basis for the expansion in \eqref{holonomy_expansion}, for instance, ${\frac{1}{2}(\mathbb{L}_0\varepsilon_{(\@p\@)}\@)^2\.\partial_{a}\bm{A}_{b}(0)\.\dot{\ell}^a_{(\@p\@)}\@(0)\.\dot{\ell}^b_{(\@p\@)}\@(0)}$. Thus, in this expansion, all the elements are projected toward the only direction determined by the path at its initial point.

Analogously, by expanding the holonomy $h_{qr}$ around the smallest loop (closing two edges, $l_q$ and $l_r$ that emanate from a single point) one finds
\begin{align}
\label{loop_holonomy_expansion}
h_{(\@q\@)\@(\@r\@)}^{\mp1}[\bm{A}]=\mathds{1}\mp\alpha\.\mathbb{L}_0^2\.\varepsilon_{(\@q\@)}\varepsilon_{(\@r\@)}\bm{F}_{(\@q\@)\@(\@r\@)}+\mathcal{O}(\varepsilon^3)\,.
\end{align}
The parameter $\alpha$ takes value $1/2$ in the case of a triangular loop and $1$ in the case of a quadrilateral one (the graph's structure depends on a selected tessellation of $M$), while $\bm{F}_{qr}:=F_{qr}^I\bm{t}_I$ is the curvature of $\bm{A}$.

In the LQG's regularization procedure \cite{Thiemann:1996aw,Thiemann:2007zz}, one is required to invert relations \eqref{holonomy_expansion}, \eqref{loop_holonomy_expansion} to replace the Ashtekar-Barbero connection and its curvature with appropriate functionals of holonomies. Both these relations, are expanded up to the same order, however, the former consists of the derivative of $\bm{A}$. Therefore, in this standard procedure, the terms of order $\varepsilon^2$ in expression \eqref{holonomy_expansion} are omitted. Although the application of these expansions is unbalanced, by expanding holonomies and taking the limit $\varepsilon\to0$ one recovers the original formula. However, finding an equivalent of \eqref{holonomy_expansion} in the form of a polynomial of $\bm{A}$, inverting it, and applying in the same order of $\varepsilon$, would provide a preferred formulation.

	%%%%%%%%%	%%%%%%%%%	%%%%%%%%%	%%%%%%%%% 

	%%%%%%%%%	%%%%%%%%%	%%%%%%%%%	%%%%%%%%%
\section{Lie algebra}\label{Sec_Algebra}

\noindent
Let $U(\bm{\theta})$ denotes a representation of a connected Lie group described by a finite set of real continuous parameters $\theta^I$ and Hermitian generators $\bm{s}_I$. Wigner's theorem \cite{Wigner:1931,Wigner:1939cj} tells that any representation of a symmetry transformation of a ray space is either a unitary and linear or else antiunitary and antilinear transformation of a Hilbert space. By expanding $U(\bm{\theta})$ around a trivial transformation, i.e. the identity, one can focus only on the unitary generators \cite{Weinberg:1995mt}. Consequently, in a finite neighborhood of the identity, one obtains the expansion
\begin{align}
\begin{split}
\label{representation_expansion}
U(\bm{\theta})=&\;\mathds{1}+\i\.\theta^I\bm{s}_I-\frac{1}{2}\theta^J\theta^K\bm{s}_J\bm{s}_K
\\
&-\frac{\i}{2}\theta^J\theta^KC^I_{\ JK}\bm{s}_I+\mathcal{O}(\theta^3)\,.
\end{split}
\end{align}
Assuming the representation of the same Lie group as in section \ref{Sec_Holonomy}, $C^I_{\ JK}$ are the same real structure constants, resulting from the following Lie bracket,
\begin{align}
\label{Lie_complex}
[\bm{s}_J,\bm{s}_K]=\i\.C^I_{\ JK}\bm{s}_I\,.
\end{align}
By comparing expressions \eqref{Lie_real} and \eqref{Lie_complex}, one finds the explicit form of the internal representation generators of $\bm{A}$,
\begin{align}
\label{generators}
\bm{t}_I=-\frac{\i}{2}\bm{s}_I\,,
\end{align}
where $\bm{s}_I$ is Hermitian and unitary.

The next natural question is whether it is possible to compare formulas \eqref{holonomy_expansion} and \eqref{representation_expansion}. Although the expanded elements appear similar, one does not have to correspond to another. Moreover, it is needed to verify if any of the terms in \eqref{holonomy_expansion} does not correspond to $\mathcal{O}(\theta^3)$. Alternatively, to verify if any of the terms in \eqref{representation_expansion} does not correspond to $\mathcal{O}(\varepsilon^3)$.

	%%%%%%%%%	%%%%%%%%%	%%%%%%%%%	%%%%%%%%%
\section{Wigner's construction\\of operators}\label{Sec_Wigner}

\noindent
LQG postulates the construction of links-located Hilbert for the Ashtekar connection operators $\hat{\bm{A}}$, which are the gauge representations of the symmetry determined by holonomies \cite{Ashtekar:2004eh,Thiemann:1996aw,Thiemann:2007zz}. This postulate requires the following identification of the group elements
\begin{align}
\label{holonomy_identification}
U(\bm{\theta})=h_{(\@p\@)}^{-1}\big[\bm{A}\big]
\end{align}

The connection coefficient $\bm{A}_{(\@p\@)}=A^I_{(\@p\@)}\bm{t}_I$, oriented along the direction of the link $l_p$, is an element of the Lie algebra in \eqref{Lie_real}. The idea to replace this gauge field with the related holonomy is inspired by the Wilson loop representation \cite{Wilson:1974sk} constructed on a piecewise linear lattice. In the case of the construction of LQG, the Hilbert spaces can be located at non-linear links. It is possible as a result of the relation in \eqref{holonomy_identification}.

The construction of Hilbert spaces at links for operator analogs of the gauge representations $\bm{\alpha}$ requires the following conditions, \textit{cf.} \cite{Wigner:1931,Wigner:1939cj}.
\\
(i) The Lie group element $U(\bm{\alpha})$ must be the exponential map from the representation $\bm{\alpha}$. Thus $U(\bm{\alpha})$ can be indicated as the group element that determines quantum symmetry transformations.
\\
(ii) In a sufficiently small neighborhood of the identity, all operators are unitary. The group has an identity element generated by the indication of the neighborhood. Consequently, $U(\bm{\alpha})$ is expandable as follows,
\begin{align}
\begin{split}
\label{Wigner_expansion}
U(\bm{\alpha})=&\;\mathds{1}-\alpha^I\bm{t}_I+\frac{1}{2}\alpha^J\alpha^K\bm{t}_J\bm{t}_K
\\
&+\frac{1}{2}\alpha^J\alpha^K[\bm{t}_J,\bm{t}_K]+\mathcal{O}(\alpha^3)\,.
\end{split}
\end{align}
\textit{cf.} \cite{Weinberg:1995mt}. It is worth mentioning that this expansion of the group element is the one, given in \eqref{representation_expansion}, adjusted to the normalization in LQG.

All holonomies are located along links. Therefore, for simplicity, to discuss a candidate for a quantum symmetry group element, only a single link-related Hilbert space construction is going to be analyzed. To follow the notation introduced in section \ref{Sec_Holonomy}, the link's length and orientation are specified by the relation
\begin{align}
\label{holonomy_specification}
\bm{\alpha}=\!\int_{\@0}^{\mathbb{L}_0\varepsilon_{(\@p\@)}}\!\!\!\!\!\!\bm{A}
=\sum_{n=0}^{\infty}\frac{1}{n!}\frac{\mathrm{d}\bm{\alpha}}{\mathrm{d}\varepsilon_{(\@p\@)}}
\@\bigg|_{\@\varepsilon_{(\@p\@)}\@=0}\varepsilon_{(\@p\@)}^n\,,
\end{align}
where $\varepsilon_{(\@p\@)}\in(0,1)$. This relation defines the connection as the candidate for the operator representation of holonomy at the quantum level in the short length limit.

The construction of the quantum representation of holonomies requires that the quantity defined in \eqref{holonomy_expansion} forms a group. It is easy to check that this condition, namely condition (i), is satisfied,
\begin{align}
\begin{split}
\label{holonomy_group}
\!\!h_{(\@p\@)}^{-1}[\bm{A}]h_{(\@p\@)}^{-1}[\bm{A'}]=&\;\mathds{1}
-\mathbb{L}_0\varepsilon_{(\@p\@)}\big(\bm{A}_{(\@p\@)}\@+\@\bm{A'}_{(\@p\@)}\big)
\\
&-\frac{1}{2}(\mathbb{L}_0\varepsilon_{(\@p\@)})^2\.\partial_{(\@p\@)}\big(\bm{A}_{(\@p\@)}\@+\@\bm{A'}_{(\@p\@)}\big)
\\
&+\frac{1}{2}(\mathbb{L}_0\varepsilon_{(\@p\@)})^2\big(\bm{A}_{(\@p\@)}\@+\@\bm{A'}_{(\@p\@)}\big)^{\@2}
\@+\mathcal{O}(\varepsilon^3)\!\!\!
\\
=&\;h_{(\@p\@)}^{-1}\big[\bm{A}_{(\@p\@)}\@+\@\bm{A'}_{(\@p\@)}\big]\,.
\end{split}
\end{align}

Condition (ii) requires first to verify the hypothesis formulated at the end of Sec.~\ref{Sec_Algebra}. The algebraic expansion in \eqref{Wigner_expansion} is required to ensure that the quantum connection is the representation of a symmetry group on the Hilbert space, which is constructed by Wigner matrices of holonomies \cite{Wigner:1931}. It is convenient to approximate the power series representation in \eqref{holonomy_specification} by its expansion up to the quadratic order, analogously to \eqref{holonomy_expansion}. The lowest order of the resulting expression
\begin{align}
\label{holonomy_specified}
\bm{\alpha}=0
+\mathbb{L}_0\varepsilon_{(\@p\@)}\bm{A}_{(\@p\@)}
+\frac{1}{2}(\mathbb{L}_0\varepsilon_{(\@p\@)}\@)^2\.\partial_{(\@p\@)}\bm{A}_{(\@p\@)}
+\mathcal{O}(\varepsilon^3)
\end{align}
is linear in $\varepsilon_{(\@p\@)}$. This formula entails the following relation,
\begin{align}
\label{parameters}
\mathcal{O}(\alpha^3)=\mathcal{O}(\varepsilon^3)\,.
\end{align}
Consequently, the right hand-sides of equations \eqref{holonomy_expansion} and \eqref{Wigner_expansion} can be identified.

Directly identifying all pairs of the elements of the expansions in \eqref{holonomy_expansion} and \eqref{Wigner_expansion}, one finds the following identity,
\begin{align}
\label{Cartan}
\alpha^J\alpha^K[\bm{t}_J,\bm{t}_K]
=(\mathbb{L}_0\varepsilon_{(\@p\@)}\@)^2\big[\bm{A}_{(\@p\@)},\bm{A}_{(\@p\@)}\big]+\mathcal{O}(\varepsilon^3)=0\,.
\end{align}
This equation expresses the Lie brackets of symmetric quantities, namely the pair of equally oriented gauge fields. Thus, it indeed vanishes. This result demonstrates that the holonomy expansion in \eqref{holonomy_expansion} coincides with the group expansion around identity. It also confirms the internal consistency of the projection in \eqref{holonomy_expansion} on a single spatial direction.

	%%%%%%%%%	%%%%%%%%%	%%%%%%%%%	%%%%%%%%%
\section{Conclusions}\label{Sec_Conclusions}

\noindent
By demonstrating that the holonomy expansion around a short length of links equals the expansion around identity, proofs that the links-located Hilbert spaces in LQG satisfy Wigner's theorem \cite{Wigner:1931,Wigner:1939cj}. Thus, this quantum theory on a lattice can be defined by the standard Heisenberg-DeWitt \cite{Heisenberg:1929xj,DeWitt:1967yk} construction of operators, known as the canonical representation (in quantum mechanics).

Moreover, the equality of the analyzed expansions also demonstrates the reality of Ashtekar variables. The one around unity in \eqref{representation_expansion} and \eqref{Wigner_expansion} is the expansion of a unitary operator. Therefore, if the Ashtekar connection was imaginary, the related parameter would also need to be imaginary, $\varepsilon_{(\@p\@)}\in(0,\i)$. The complex connection, which would be composed of real and imaginary terms, is excluded. The imaginary Ashtekar connection would also lead to the imaginary length of the graph's edges, and in consequence, to a non-physical model.

Considering then the $\mathfrak{su}(2)$ representation, the set of the Hermitian and unitary generators, $\bm{s}_I=\bm{\sigma}_I$, is composed of the Pauli matrices, and the structure constants $C^I_{\ JK}=2\epsilon^I_{\ JK}$ are the totally antisymmetric Levi-Civita tensor coefficients. This is the standard choice in LQG. In this case, the $\mathfrak{su}(2)$-valued Ashtekar-Barbero connection becomes a canonically conjugate partner to the densitized dreibein. Together they form the real Ashtekar variables, thus allowing the variables to represent the Holst Hamiltonian \cite{Holst:1995pc} in a quantizable form after the lattice regularization.

Finally, it is worth mentioning that the same precision, as the one in the analysis in this article, should be postulated in the construction of the holonomy formulation of the lattice-smeared Hamiltonian of LQG. Namely, one should use the expansions up to the quadratic order in \eqref{holonomy_expansion}, as it is postulated concerning the relation in \eqref{loop_holonomy_expansion} in LQG. This precision will ensure that the symmetry transformations described by the Hamiltonian on a lattice are implemented rigorously. The standard holonomy smearing in LQG is not as precise as in the construction of the related Hilbert spaces \cite{Ashtekar:2004eh,Thiemann:1996aw,Thiemann:2007zz}. There are two methods to increase the symmetry-preserving in the lattice formulation of the Hamiltonian. One can symmetrize the locations of connections around links-related holonomies \cite{Bilski:2021_RCT_II,Bilski:2020poi}. Alternatively, one can select the connections' distributions $\bm{\alpha}$ in \eqref{holonomy_specification} as the canonical quantum fields \cite{Bilski:2021_RCT_II,Bilski:2021ysc}. It is also worth noting that in cosmology, the connections are spatially constant \cite{Bilski:2019tji}. Hence, both methods lead to the same results \cite{Bilski:2021fki}.

	%%%%%%%%%	%%%%%%%%%	%%%%%%%%%	%%%%%%%%%

	%%%%%%%%%	%%%%%%%%%	%%%%%%%%%	%%%%%%%%%

{\acknowledgments
\noindent
This work was partially supported by the National Natural Science Foundation of China grants Nos. 11675145 and 11975203.
The author thanks Piotr Latasiewicz for language editing in the first version of the manuscript.}

	%%%%%%%%%	%%%%%%%%%	%%%%%%%%%	%%%%%%%%%

	%%%%%%%%%	%%%%%%%%%	%%%%%%%%%	%%%%%%%%%

\end{document}